\documentclass[superscriptaddress,twocolumn]{revtex4-2}
\usepackage{graphicx,bm,dsfont,amsmath, amsthm, amssymb,mathrsfs}
\usepackage{setspace}
\usepackage{orcidlink}
\usepackage[utf8]{inputenc}
\usepackage{hyperref}
\usepackage{mdframed}
\usepackage{xcolor}
\usepackage{soul}

\begin{document}

\title{Comment on ``Experimentally adjudicating between different causal accounts of Bell-inequality violations via statistical model selection''}
\author{Jonte R.\ Hance\,\orcidlink{0000-0001-8587-7618}}
\email{jonte.hance@bristol.ac.uk}
\affiliation{Graduate School of Advanced Science and Engineering, Hiroshima University, Kagamiyama 1-3-1, Higashi Hiroshima 739-8530, Japan}
\affiliation{Quantum Engineering Technology Laboratories, Department of Electrical and Electronic Engineering, University of Bristol, Woodland Road, Bristol, BS8 1US, UK}
\author{Sabine\ Hossenfelder\,\orcidlink{0000-0002-2515-3842}}
\affiliation{Munich Center for Mathematical Philosophy, Ludwig-Maximilians-Universit\"at M\"unchen, Geschwister-Scholl-Platz 1,
D-80539 Munich, Germany}
\date{\today}

\begin{abstract}
In a recent paper (Phys. Rev. A \textbf{105}, 042220 (2022)), Daley \emph{et al} claim that some superdeterministic models are disfavoured against standard quantum mechanics, because such models overfit the statistics of a Bell-type experiment which the authors conducted. We add to the discussion by providing additional context about how few superdeterministic models fall into the category they analyse, and by emphasising that overfitting, while better as a measure of finetuning than other measures given in the literature, does not necessarily indicate a model is universally bad.
\end{abstract}

\maketitle

\section{From finetuning to overfitting}

In a recent paper \cite{Daley2022Adjudicating}, Daley \emph{et al} claim it is possible to show that a class of superdeterministic models are disfavoured against standard quantum mechanics, because they overfit the statistics of Bell-type experiments.
Below we will comment on their argument. However, we would like to start by explaining why their recent analysis is great progress. 

First, by using overfitting and standard definitions for explanatory power, the authors avoid the metaphysical notions of `finetuning' that have stalled progress in the foundations of physics for decades. 

The only scientifically relevant notion of finetuning is overfitting. All other notions of finetuning are properties that a model may or may not have, but which do not a priori imply the model has a problem. However, metaphysical notions of finetuning are unfortunately in wide use: both in cosmology and particle physics \cite{Hossenfelder2021Screams}, and through the previously-used notion of `faithfulness' \cite{Wood2015FineTuning} in quantum foundations. 

In all those cases, it well may be that a model has such a property, but so long as the model explains data well (has explanatory power), such a property is not a valid reason to discard the model. In contrast, a model that overfits experimental data and has low explanatory power is scientifically untenable, not to mention useless.
(We note in passing that both General Relativity and the Standard Model of Particle Physics are `finetuned' according to some metaphysical definitions, so clearly such a property does not signal a lack of explanatory power.)

A second reason the new paper is a step forward is that the authors acknowledge that the class of superdeterministic models is wider than the {set} they {analyse using their method}. That is, their discussion is considerably more nuanced than previous studies of superdeterminism, whose major purpose seems to have been to confirm the firmly-held belief that it is some kind of unscientific conspiracy theory. However, we will next explain why all superdeterministic models present in the literature are those that Daley \emph{et al} did not analyse using their method.

\section{Superdeterminism is not classical}

Daley \emph{et al} state that they {analyse} classical superdeterministic models {using their method}. For justification, the reader is referred to Wood and Spekkens's paper \cite{Wood2015FineTuning} --- however, \cite{Wood2015FineTuning} does not provide any such justification.

Superdeterministic models are in general not classical in any meaningful sense of the word. All examples we know of (Brans's model \cite{Brans1988Model}; Palmer's invariant set theory \cite{Palmer2020Discretization}; t'Hooft's cellular automaton model \cite{tHooft2016Cellular}; Ciepielewski, Okon and Sudarsky's model \cite{Ciepielewski2020Superdeterministic}; and Donadi and Hossenfelder's toy model \cite{Donadi2020SuperdetToy}) use wave-functions and density matrices, superpositions and entanglement, just like quantum mechanics.

In a superdeterministic model, these quantities describe an ensemble \cite{Hance2021Ensemble} rather than an ontic state (hence rendering the measurement update of the wavefunction purely epistemic), but that doesn't make superdeterministic models classical. This should not be surprising, given the purpose of superdeterminism is not to return to classical mechanics, but merely to return to locality. 

It is somewhat unclear why Daley \emph{et al} would focus their analysis on models with this property, given that it is ostensibly not fulfilled in most models presented in the literature.

\section{Physics beyond quantum mechanics}

Another problem with Daley \emph{et al}'s argument is that, by selecting these classical superdeterministic models as the ones most worth analysing, and by analysing them with respect to overfitting a Bell-type experiment, they potentially misrepresent the very purpose of the superdeterministic models present in the literature.

The key property of a superdeterministic model is that it is a locally causal hidden variables model which violates statistical independence: that is, the probability distribution of the hidden variables is correlated with the detector settings at the time of measurement \cite{Hossenfelder2020Rethinking,Hossenfelder2020SuperdeterminismGuide}. When averaged over the hidden variables, a superdeterministic model must reproduce quantum mechanics to agree with experimental data.

As such, superdeterministic models are candidate models for a more fundamental physics that underlies quantum mechanics. They describe ``phenomenology beyond quantum mechanics'' \cite{adlam2023taxonomy,hossenfelder2023quantum} in much the way that we have models for phenomenology beyond the standard model (such as supersymmetry \cite{haber1985search}). 

The difference between these two phenomenologies is the parameter range in which we expect the deviations to become relevant. Physics beyond the standard model is expected to become relevant at high energies, or in experiments sensitive to rare interactions and weak coupling. Physics beyond quantum mechanics, on the other hand, should appear either along the Heisenberg cut, with some parameter that quantifies the transition from quantum to classical behaviour (loosely, when objects become sufficiently macroscopic in size), or when dealing with measurements far faster than those done in Daley et al's experiment (to probe super-quantum correlations between measurements, with coherence times too short to have been seen yet \cite{hossenfelder2011testing}). In contrast to this latter limit, we shall in the following refer to the domain in which Daley et al are measuring as the ``large-$t$ limit'' of superdeterminism, which reproduces quantum mechanics. We unfortunately cannot exactly state what the limit, is because we don't know exactly under which circumstances the new physics appears --- this needs to be experimentally probed.  

Even though we do not know the exact parameter range in which deviations from quantum mechanics should occur, we can illustrate the role of superdeterministic models in a comparison to statistical mechanics. In this analogy, superdeterministic models relate to quantum mechanics similarly to how statistical mechanics relates to thermodynamics. 

Using this analogy let us try to ask a similar question as \cite{Daley2022Adjudicating} ask of quantum mechanics. Suppose you want to know whether thermodynamics is fundamental or emergent from the underlying ``hidden variables" theory that is atomic physics. You take an observable like, say, the equilibrium temperature of a two-component system. That is, you have two fluids, mix them together, wait until they reach equilibrium, and ask what the final temperature is. The equation for this in equilibrium thermodynamics is
\begin{equation}
    T_f = \frac{c_1 m_1 T_1 + c_2 m_2 T_2}{c_1 m_1+c_2 m_2} \label{Eq:1}
\end{equation}
where $T_f$ is the final temperature of the mixture, and $c_i$, $m_i$ and $T_i$ are the specific heat capacity, mass, and temperature of component $i$ (for $i=1,2$).

However, for statistical thermodynamics,
\begin{equation}
\begin{split}
        &T_f=\left( \frac{d}{dE}k_B \ln{(W_f)}\right)^{-1},\; \text{where}\\
        &W=\int...\int \frac{1}{h^nC}f(\frac{H-E}{\omega})dp_1...dp_ndq_1...dq_n
\end{split}
\end{equation}
where $k_B$ is Boltzmann's constant, $W$ is the number of microstates the system can be in, across a $2n$-dimensional phase space (i.e. possessing $n$ pairs of conjugate coordinates $p_i,q_i$), $H$ is the system Hamiltonian, $E$ is the total system energy (as specified by the sum of the energies of the two mixed fluids), $f$ is a function of these with width $\omega$, $h$ is a dimension-setting constant in units of energy-time, and $C$ is a correction factor due to overcounting \cite{landau_statistical_1980}.

If all you want to do is fit Eq.~\ref{Eq:1}, then having multiple free parameters for each single atom in the fluid is clearly unnecessary. If you have that many parameters, they will fit every single fluctuation, however minor --- you can make it fit everything. It's bad science, it's overfitting, and it's a terrible model. In terms of explanatory power, equilibrium thermodynamics is much preferable. 

But of course that argument makes no sense, because if you are asking for the prediction for a statistical average, you need a distribution for the hidden variables (that is, the position and momenta of atoms), and then you need to calculate the averages you obtain from this. The distribution is part of the model. In the large-$N$ limit, deviations from the average become increasingly unlikely, hence equilibrium thermodynamics is the most economic explanation you can have.  

To come back to superdeterminism, the average of the large-$t$ limit of a superdeterministic theory just gives quantum mechanics. This means whenever the system is large and/or warm, or noisy, and measurements are slow, etc (compared in scale to those suggested in \cite{hossenfelder2011testing}), the quantum mechanical prediction is the one with the highest explanatory power. In that limit, we cannot test superdeterminism with violations of Bell's inequality \cite{Hossenfelder2020Rethinking,Hossenfelder2020SuperdeterminismGuide}, and neither can we test it with any other bound or inequality---it will give the same answer as quantum mechanics. So how can Daley \emph{et al} find superdeterminism to be any worse than quantum mechanics?

It's because, from a purely axiomatic perspective, the distribution of the hidden variables in the large-$t$ limit of superdeterminism must be logically equivalent to Born's rule. One can always trivially reverse-engineer the distribution of the hidden variables from the requirement that the result gives Born's rule. If one then uses Born's rule to fit data with quantum mechanics, but does not use the corresponding distribution for the hidden variables of the superdeterministic model, then the superdeterministic model, when trained on data, will overfit. By the same logic though, we would have to rule out the atomic hypothesis, because it is parametrically wasteful if we just want to find a model that fits equilibrium temperatures.

Can it happen in statistical mechanics that the temperatures actually evolve away from the equilibrium value? Yes. But in a large system, it is extremely unlikely, so it is likely you never see it. Can it happen in a superdeterministic theory that you see a deviation from Born's rule? Yes, but for a comparably slow system (as Daley \emph{et al} consider), you will for all practical purposes never see it. The situation for superdeterminism is hence pretty much the same as in Bohmian mechanics with the equilibrium hypothesis, a case which Daley \emph{et al} do discuss. As long as the equilibrium hypothesis is fulfilled, Bohmian mechanics is parameterically restricted and reproduces quantum mechanics. It's the same with superdeterminism in the large-$t$ limit. (This is something Daley et al themselves do admit.)

Daley \emph{et al}'s argument is right in this sense: you do not need the flexibility of having possibly different initial distributions if the only observable you measure is the equilibrium average, for which we already know Born's rule is the best you can do. It is not like anyone claims that quantum mechanics is a bad theory. It is a great theory in certain limits (just like thermodynamics is). This is exactly why you have to look at a system which is \textbf{outside} that limit. 

If we are outside the limit where deviations from quantum mechanics are negligible, then Daley \emph{et al} are correct that one can in principle test superdeterminism even with Bell-type experiments. This is contrary to what is often stated in the literature: that the superdeterminism loophole can't be closed (though the opposite has also been claimed \cite{abellan2018challenging}). However, being able to see such deviations in certain circumstances has nothing to do with violations of Bell's inequality. These would still be violated. Rather, one would see deviations from Born's rule---in particular, one might expect a time-correlation of measurement outcomes \cite{hossenfelder2011testing}. There are easier ways to test this than with Bell-type experiments. That is to say if one takes testing superdeterminism with Bell-type experiments to mean that one establishes violations of Bell's inequality, then it is correct that the superdeterminism loophole cannot be closed. 

\section{Conclusion}

We have argued that when considering Daley \emph{et al}'s claim that superdeterministic models are disfavoured against standard quantum mechanics, because such models overfit the statistics of a Bell-type experiment, we need to remember that overfitting for a given scenario is not necessarily bad in all situations, and given superdeterministic models are often used to try to consider physics beyond quantum mechanics, we should expect them to overfit a test such as the one Daley \emph{et al} perform.

We believe it would be a good idea if discussions about properties of superdeterministic models considered at least one superdeterministic model from the literature. There are sufficiently many of these, and analysing them directly, with reference to their aims and domain of applicability, seems the best route for determining whether or not superdeterministic models are fine-tuned.

\textit{Acknowledgements}
We thank Tim Palmer for useful discussions. SH acknowledges support by the Deutsche Forschungsgemeinschaft (DFG, German Research Foundation) under grant number HO 2601/8-1. JRH acknowledges support from Hiroshima University's Phoenix Postdoctoral Fellowship for Research, and the University of York's EPSRC DTP grant EP/R513386/1. JRH and JR acknowledge support from the UK Quantum Communications Hub, funded by the EPSRC grants EP/M013472/1 and EP/T001011/1. 

\bibliographystyle{unsrturl}
\bibliography{ref.bib}

\end{document}